# Direct generation of intense extreme ultraviolet supercontinuum with 35 fs, 11 mJ pulses from a femtosecond laser amplifier

Bin Zeng[1,2], Wei Chu[1,2], Guihua Li[1,2], Jinping Yao[1], Jielei Ni[1,2], Haisu Zhang[1,2], Ya Cheng[1,§], and Zhizhan Xu[1,†], Yi Wu[3], Zenghu Chang[3]

[1]*State Key Laboratory of High Field Laser Physics, Shanghai Institute of Optics and Fine Mechanics, Chinese Academy of Sciences, Shanghai 201800, China*

[2]*Graduate School of Chinese Academy of Sciences, Beijing 100039, China*

[3]*CREOL and Department of Physics, University of Central Florida, Orlando, Florida 32816, USA*

[§]*Email: ya.cheng@siom.ac.cn*

[†]*Email: zzxu@mail.shcnc.ac.cn*

**Abstract:**

We report on the generation of intense extreme ultraviolet (EUV) supercontinuum with photon energies spanning from 35 eV to 50 eV (i. e., supporting an isolated attosecond pulse with a duration of ~271 as) by loosely focusing 35 fs, 11 mJ pulses from a femtosecond laser amplifier into a 10-mm long gas cell filled with krypton gas. The dramatic change of spectral and temporal properties of the driver pulses after passing through the gas cell indicates that propagation effects play a significant role in promoting the generation of the EUV supercontinuum.

The successful generation of attosecond pulses has enabled time resolved investigation of light-matter interaction at the natural time scale of electrons [1]. To date, high-order harmonic generation (HHG) still remains the standard workhorse for generation of isolated attosecond pulses (IAPs) [2], which requires the use of either externally compressed carrier-envelop-phase (CEP) controlled few-cycle pulses [3] or waveform-synthesized multi-cycle pulses [4, 5]. The employment of sophisticated waveform tailoring and control techniques often leads to increased complexity of experimental setup, limited driver pulse energies, and extra optical losses of the driver pulses. These obstacles hamper the widespread use of attosecond technology for revealing ultrafast dynamics in atomic, molecular, and condensed matter systems. Therefore, new techniques that can greatly simplify the attosecond pulse generation and make a more efficient use of the driver pulses are highly in demand.

Recently, it has been shown that IAPs can be directly generated from a filament formed by loosely focusing the driver laser beam into a semi-infinite gas cell filled with argon at atmospheric pressure [6]. In this case, the multi-cycle driver pulses with a pulse energy of ~1 mJ and an initial pulse duration of ~35 fs were automatically transformed to pulses with complex temporal structures during the filamentation process. In particular, intensity spikes with a duration approaching single optical cycle could emerge in the transformed pulses within which IAPs were produced. Obviously, this technique can greatly simplify the attosecond pulse generation as it does not have any specific requirements on the initial waveform of driver pulses except the control of CEP. Recently, when we carry out HHG experiments with high-energy driver pulses aiming at producing high-brightness X-ray radiation (i. e., the original purpose of this experiment is to generate intense high-order harmonics but not supercontinuum), surprisingly, we observed that EUV supercontinuum emerges when Fourier-transform-limited ~35 fs, 11-mJ femtosecond laser pulses were loosely focused into a 10-mm long gas cell filled with low-density krypton gas. It is noteworthy that although our gas cell was only 10-mm-long, there were two holes drilled by the driver laser pulses at the two ends of the cell through which leakage of krypton gas inevitably occurs. The diameter of the holes was measured to be ~2 mm. As we will show below, the leaking gas from the holes could form a favourable density distribution of the krypton gas outside the gas cell for inducing filamentation and self-compression of the driver pulses, facilitating generation of IAPs. In addition, we find that control of the gas pressure, driver pulse energy, and the cell length is crucial in this experiment. With these optimizations, we obtain single-shot spectra of intense EUV supercontinua covering the spectral range from 35 eV to 50 eV.

The experimental setup is sketched in Fig. 1, which is exactly the same one used in our previous HHG experiments without any modification [7, 8]. Although the previous experiments were focused on the HHG driven by long wavelengths, in this experiment, an 800 nm source was directly employed for HHG to utilize its full pulse energy. The laser system (Legend Elite Cryo PA, Coherent, Int.) delivers ~35 fs (FWHM) pluses with a central wavelength at 800 nm and a single pulse energy of ~11 mJ. The beam diameter was measured to be ~8.8 mm ($1/e^2$). To generate Fourier-transform-limited driver pulses with shortest durations, we adjusted the distance between the grating pairs in the compressor of the femtosecond laser amplifier. A 10-mm-long gas cell filled with krypton gas was mounted in the vacuum chamber for generating high-order harmonics. The gas pressure can be adjusted in the range from 0 mbar to 200 mbar for optimization of the HHG process. The back pressure inside the vacuum chamber was ~$10^{-3}$ mbar. A flat-field grating spectrometer equipped with a soft-x-ray CCD (Princeton Instruments, 1340×400 imaging array PI:SX 400) was used to characterize the harmonic emission. We employed a 150 nm thick aluminium foil to block the lower-order harmonics and the residual infrared pulses. Since the driver pulses were not CEP stabilized, all the harmonic spectra presented here were captured in single shots. Although our previous experiments were mostly carried out by focusing ~1-2 mJ, ~35 fs pulses with lenses of focal lengths ranging from 15 cm to 40 cm, in this experiment a significant higher pulse energy of ~11 mJ (i. e., the maximum pulse energy affordable by our amplifier) was used, and a loose focusing scheme with a 100-cm-focal-length was employed. Under these conditions we observed that filamentation occurred at gas pressures above ~30 mbar, as indicated by the arrows in the inset of Fig. 1. A long filament channel can be clearly seen at the both sides of the gas cell.

Fig. 2(a) shows a typical two-dimensional (2D) single-shot HHG spectrum observed by focusing the Fourier-transform-limited ~35fs, 11-mJ pulses into the 10-mm-long gas cell filled with ~30 mbar krypton gas. Dramatically, we observe a supercontinuum covering the spectral range from ~35 eV to ~50 eV (signals below ~35 eV were not recorded due to limited size of X-ray CCD). Fig. 2(b) presents the EUV supercontinuum spectrum for the signal distributed on the black dashed line in Fig. 2(a) (i. e., the divergence angle of 0 mrad). By assuming that the phase of the EUV supercontinuum in Fig. 2(b) is flat, a ~271 as IAP with a clean temporal profile can be generated as shown in Fig. 2(c). The intensity of satellite pulses is well below 3% of that of the main pulse.

We found that the generation of EUV supercontinuum is sensitive to the CEP value of the driver pulses. Figs. 3(a), (b) and (c) presents HHG spectra recorded in other single shots which preserves all the same experimental conditions as used for generation of the supercontinuum spectrum in Fig. 2(a) except the CEP values, as the CEP value of our femtosecond laser amplifier was not stabilized. The HHG spectra in Figs. 3(b) and (c) show strong periodic modulation, indicating that control of CEP is still necessary in generation of IAPs with this new approach. Nevertheless, CEP-stabilized amplifiers with pulse energies up to tens of millijouls are already commercially available and used world-widely for HHG experiments, making it possible for energy scaling up in single attosecond pulse generation using this technique. The driving pulse energy is also critical for the generation of the supercontinuum. Fig. 3(d) shows a HHG spectrum recoded at a lower input energy of 8 mJ. We can clearly see that the supercontinuum disappears, probably because of the weakened nonlinear effects caused by the lower pulse energy. Moreover, the HHG spectrum is also sensitive to the length of gas cell. It can be seen in Fig. 3(e) that when shorter gas cells (e. g., 2-mm-long) is used, only discrete harmonics can be observed. This clearly indicates the important role of the propagation effects.

It is well known that generation of IAPs requires creation of a narrow time-window within which the well-known HHG process occurs only once [1, 2]. Thus, from a single-atom point of view, it should be impossible to produce the EUV supercontinuum with the multi-cycle ~35 fs driver pulses as shown in Fig. 2(a). However, HHG is also a highly nonlinear process extremely sensitive to propagation effects. As our driver pulses are very intense, filamentation can even occur at a low gas pressure of only ~30 mbar as evidenced in the inset of Fig. 1, leading to the spatio-temporal change of the driver pulses during the course of HHG. Fig. 4(a) shows that the spectrum of the driver pulses coming out from the HHG chamber is significantly broader than that of the input pulses. Moreover, the raw SHG FROG trace of the outgoing driver pulses is shown in Fig. 4(b), and the retrieved pulses are displayed in Fig. 4(d). Fig. 4(d) shows that the driver pulses were split into several intensity spikes. We would like to stress that the retrieved pulse shown in Fig. 4(d) represents the temporal profile of the pulse exiting from the end of the filament, which is not necessarily to be the same as the driver pulse responsible for the generation of the EUV supercontinuum. For example, the driver pulse may only exist in the middle portion of the filament. However, the strong modulation of the driver pulses (Fig. 4(d)) provides a clear evidence that they have undergone dramatic spectral/temporal change during the propagation through the gas cell. Thus, there is no doubt that the observed EUV supercontinuum must, at least partially, be generated due to some propagation-induced effects such as phase matching and/or ionization gating [9, 10]. Particularly, it should be noticed that in this high-energy regime, the driver pulses themselves can undergo dramatic changes during the propagation because of strong plasma generation. Therefore, for fully clarifying the underlying mechanism, sophisticated 3D simulation tools including the propagation effects are needed [11, 12] which will be our future direction.

Besides being easy to operate, the use of the long gas cell with two holes drilled at its two ends by the driver laser pulses has several other advantages for both filamentation and HHG. First of all, unlike a gas jet which usually forms a decreasing gas density in the direction perpendicular to the propagation direction of the driver laser, the gas cell with holes at its two ends can form an axially symmetrical density distribution for the driver pulses. In addition, since the holes are drilled by the driver pulses, perfect alignment can automatically be achieved. Secondly, the spread of the gas by leaking out from the holes naturally forms a favourable density gradient for single filamentation at high pulse energies [13], which is very useful for scaling up the HHG yield by use of high-energy driver pulses. Thirdly, both the length of the gas cell and the gas pressure can be easily tuned, which facilitates optimization of both propagation of driver pulse and phase matching of HHG.

In conclusion, we have successfully demonstrated generation of EUV supercontinuum with Fourier-transform-limited ~35 fs, 11 mJ driver pulses directly from an amplifier. Our experiment is carried out in low-pressure noble gas with a conventional HHG experimental setup, which could lead to great simplification of the technology for attosecond pulse generation. With higher pulse energies, this technique also holds promise for generation of intense IAPs using longer-focal-length lenses. Both these advantages will have important implementations for attosecond science and technology.

The work is supported by National Basic Research Program of China (Grant 2011CB808102), and NSFC (Grant Nos. 11134010, 10974213 and 60825406).

**References**

[1] F. Krausz, and M. Ivanov, Rev. Mod. Phys. **81**, 163 (2009).

[2] P. B. Corkum, Phys. Rev. Lett. **71**, 1994 (1993).

[3] E. Goulielmakis *et al*, Science **320**, 1614 (2008).

[4] G. Sansone *et al*, Science **314**, 443 (2006).

[5] H. Mashiko *et al*, Phys. Rev. Lett. **100**, 103906 (2008).

[6] D. S. Steingrube *et al*, New J. Phys. **13**, 043022 (2011).

[7] H. Xiong *et al*, Opt. Lett. **34**, 1747 (2009).

[8] H. Xu *et al*, Opt. Lett. **35**, 472 (2010).

[9] Z. Chang *et al*, Phys. Rev. A **58**, R30 (1998).

[10] T. Pfeifer *et al*, Opt. Express **15**, 17120 (2007).

[11] V. Tosa, H. T. Kim, I. J. Kim and C. H. Nam, Phys. Rev. A **71**, 063807 (2005).

[12] A. Couairon, H. S. Chakraborty and M. B. Gaarde, Phys. Rev. A **77**, 053814 (2008)

[13] A. Suda, M. Hatayama, K. Nagasaka and K. Midorikawa, Appl. Phys. Lett. **86**, 111116 (2005).

**Figure captions:**

Fig. 1 (Color online) Schematic of the experimental setup. The same single filament can be seen at the both sides of the gas cell as shown in the inset.

Fig. 2 (Color online) (a) 2D single-shot supercontinuum spectrum generated by the Fourier-transform-limited ~35fs, 11mJ laser pulse; (b) EUV supercontinuum spectrum for the signal distributed on the black dashed line in Fig. 2(a); and (c) temporal profile of the single attosecond pulse by inverse Fourier transform of the spectrum in (b) provided that the attochirp is fully compensated.

Fig. 3 (Color online) (a), (b) and (c) HHG spectra generated in single shots with different CEP values, giving a clear evidence on the CEP effect. (d) HHG spectrum generated at a lower energy of 8 mJ. (e) HHG spectrum generated with a 2-mm-long gas cell.

Fig. 4 (Color online) (a) Spectra of the original driver pulses (solid line) and the driver pulses coming out from the HHG chamber (dashed line). (b) Measured and (c) Retrieved FROG traces of the driver pulses coming out from the gas cell. (d) Retrieved temporal intensity (solid line) and phase (dashed line) of the driver pulses.

Fig. 1

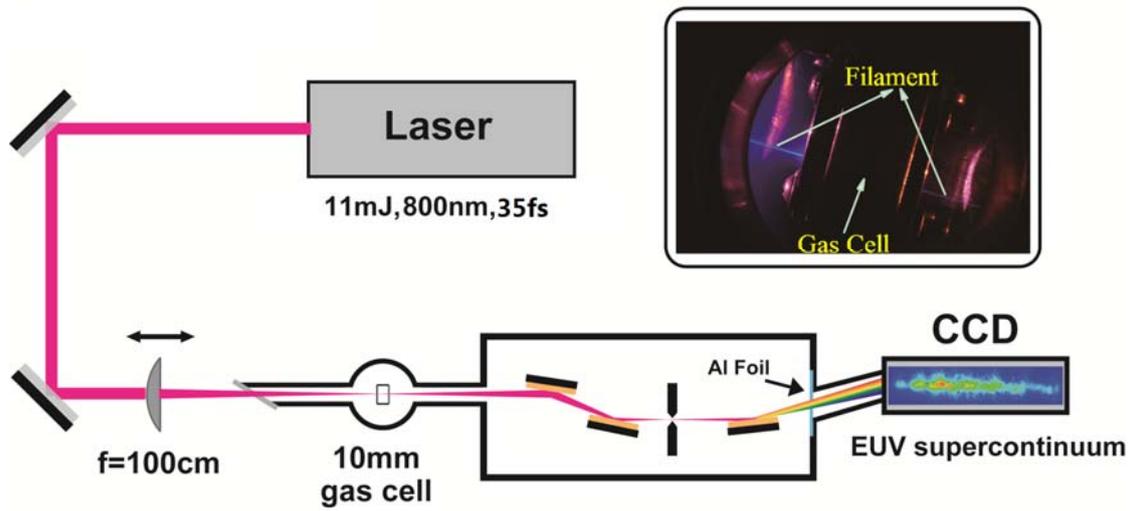

Fig. 2

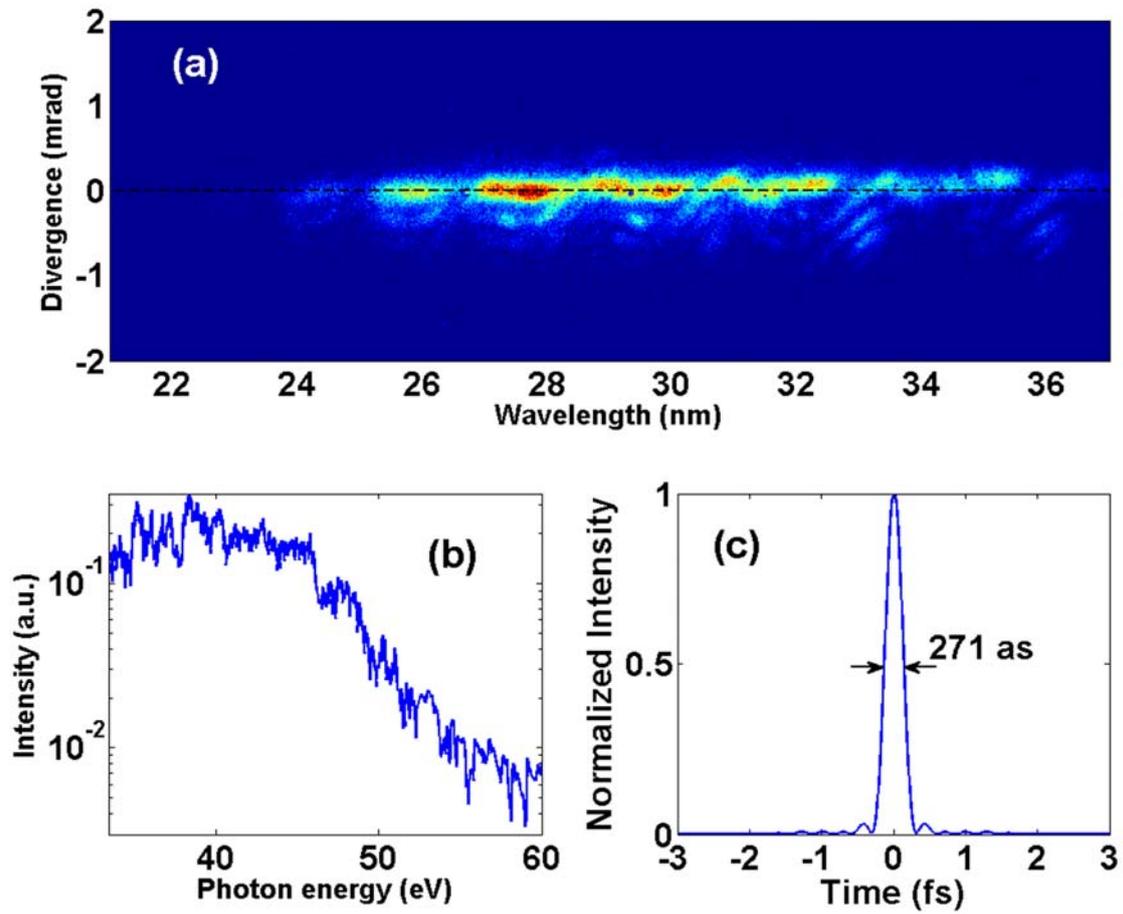

Fig. 3

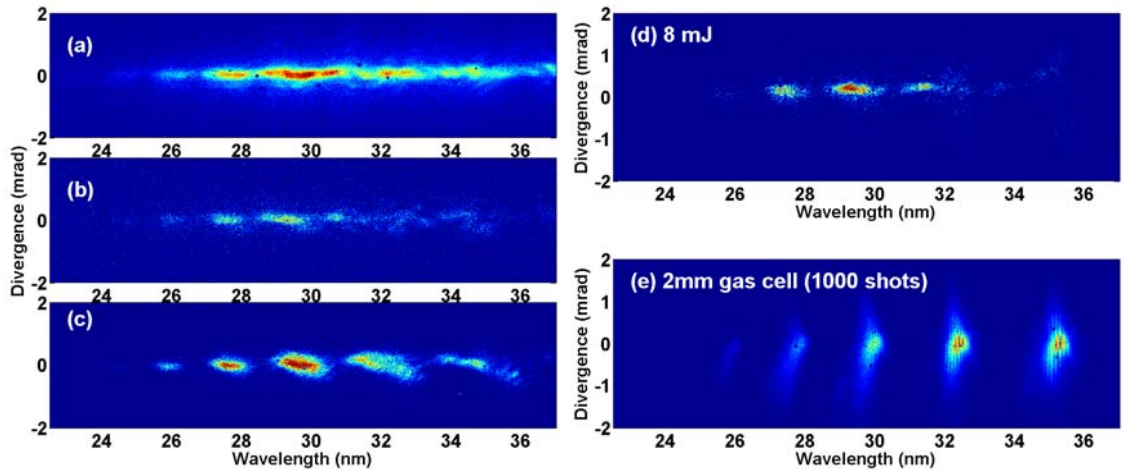

Fig. 4

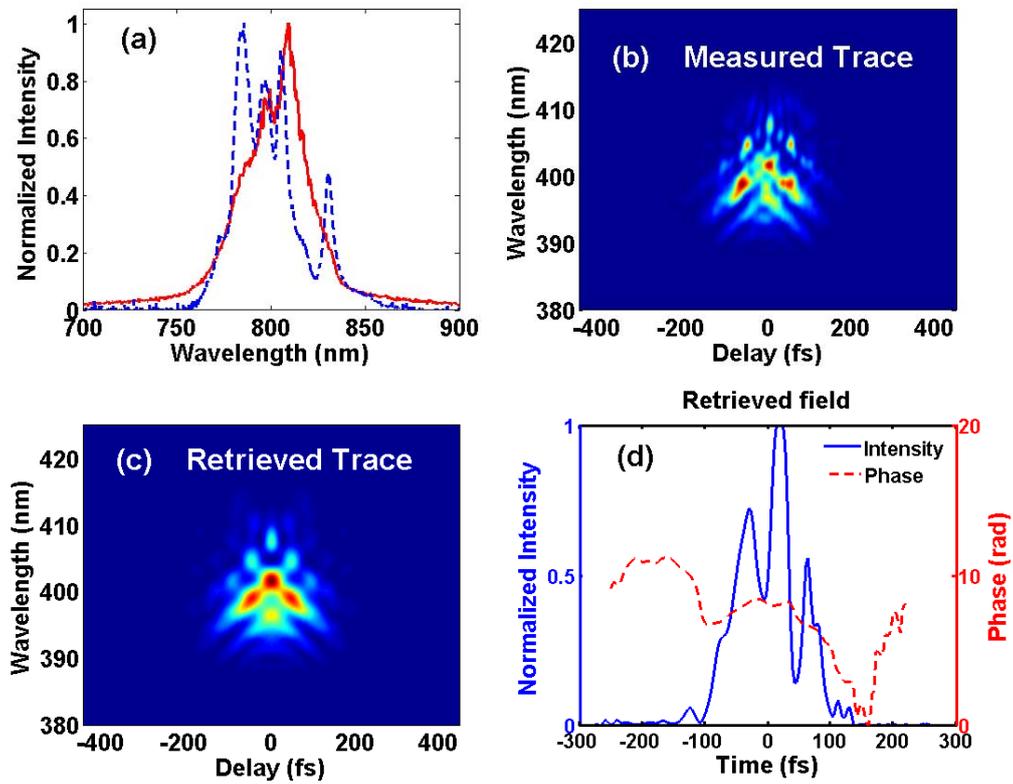